\documentclass[
12pt,
superscriptaddress,
preprint,
showpacs,
 amsmath,amssymb,
 aps,
prl,
]{revtex4}

\usepackage{graphicx}% Include figure files
\usepackage{dcolumn}% Align table columns on decimal point
\usepackage{bm}% bold math
\usepackage{amsmath}
\usepackage{color}

\usepackage{mathptmx} %HCM use times fonts (also better spacing)

\begin{document}

\break

\author{F.~C.~Niestemski}
\thanks{These authors contributed equally to this work.}
\affiliation{Stanford Institute for Materials and Energy Sciences, SLAC National Accelerator Laboratory, Menlo Park, California 94025, USA}
\affiliation{Department of Physics, Stanford University, Stanford, California 94305, USA}

\author{S.~Johnston}
\thanks{These authors contributed equally to this work.}
\affiliation{Stanford Institute for Materials and Energy Sciences, SLAC National Accelerator Laboratory, Menlo Park, California 94025, USA}
\affiliation{Institute for Theoretical Solid State Physics, IFW Dresden, Helmholtzstrasse 20, 01069 Dresden, Germany}

\author{A.~W.~Contryman}
\affiliation{Stanford Institute for Materials and Energy Sciences, SLAC National Accelerator Laboratory, Menlo Park, California 94025, USA}
\affiliation{Department of Applied Physics, Stanford University, Stanford, California 94305, USA}

\author{C.~D.~Camp}
\affiliation{Stanford Institute for Materials and Energy Sciences, SLAC National Accelerator Laboratory, Menlo Park, California 94025, USA}
\affiliation{Department of Physics, Stanford University, Stanford, California 94305, USA}

\author{T.~P.~Devereaux}
\affiliation{Stanford Institute for Materials and Energy Sciences, SLAC National Accelerator Laboratory, Menlo Park, California 94025, USA}
\affiliation{Geballe Laboratory for Advanced Materials, Stanford University, Stanford, California 94305, USA\vspace{1cm}}

\author{H.~C.~Manoharan}
\email{manoharan@stanford.edu}
\affiliation{Stanford Institute for Materials and Energy Sciences, SLAC National Accelerator Laboratory, Menlo Park, California 94025, USA}
\affiliation{Department of Physics, Stanford University, Stanford, California 94305, USA}
\affiliation{Geballe Laboratory for Advanced Materials, Stanford University, Stanford, California 94305, USA\vspace{1cm}}

%\title{Local Spectral Inversion and Bosonic Extraction of the Archetypal  Elemental Superconductor with S-Vacuum-S Tunneling Spectroscopy}
\title{Local Spectral Inversion and Bosonic Fine Structure Extraction \break
via Superconducting Scanning Tunneling Spectroscopy\vspace{0.5cm}}

\begin{abstract}
\vspace{0.7cm}
We perform the scanning tunneling spectroscopy based superconductor-vacuum-superconductor analogue to the
seminal McMillan and Rowell superconductor-insulator-superconductor device study of phonons in the archetypal elemental superconductor Pb [W.\ L.\ McMillan and J.\ M.\ Rowell, 
Phys.\ Rev.\ Lett.\ \textbf{14}, 108 (1965)].  We invert this spectroscopic data
utilizing strong-coupling Eliashberg theory to obtain a local $\alpha^2
F(\omega)$ and find broad underlying agreement with the pioneering results, highlighted by
previously unobserved electron-hole asymmetries and new fine structure which
we discuss in terms of both conventional and unconventional superconducting
bosonics.  \end{abstract}

% insert suggested keywords - APS authors don't need to do this\keywords{}

\maketitle
	
        It has been well established that the Cooper pairs of conventional
superconductors are bound by a ``pairing glue'' mediated by 
phonons \cite{PhysRevLett.14.108,BCS1957}.  The verification of 
phonon-mediated pairing was facilitated by a combination of experimental evidence
including an isotope effect study on elemental Hg \cite{PhysRev.78.487},
tunneling studies on Pb devices by McMillan and Rowell (M\&R) 
and Giaever \textit{et al.} \cite{PhysRevLett.14.108, giaever1961}, and  
phonon measurements on Pb by Brockhouse \textit {et al.} \cite{PhysRev.128.1099}. 
Since tunneling studies played such a crucial role in the original
identification of the pairing glue, point contact and device tunneling methods
have been in continual employment to study the bosonic structure of both
conventional and unconventional superconductors alike
\cite{PhysRevB.31.6096,PhysRevB.42.7953,PhysRevB.50.7177}. Additionally, with
recent technological improvements in scanning tunneling spectroscopy (STS)
experiments,  bosonic information has been recorded for various high transition
temperature (high-$T_c$) superconducting systems utilizing metallic tips
\cite{Lee2006,Niestemski2007,Niestemski2011, Pasupathy2008,
PhysRevLett.103.227001, PhysRevLett.105.167005, PhysRevLett.108.227002}.  This
STS-based setup proves superior to superconductor-insulator-superconductor (S-I-S) device tunneling for its ability to
measure spectra at single atomic locations rather than large spatial averages,
in addition to STS utilizing the cleaner vacuum barrier over the device's
sandwiched insulator.
	
%%%%%%%%%%%figure%%%%%%%%%%%%%%%%%%%%%%%%%%%%%%%%%%%%%%%%%%%%%%%%%%%%%%%%%%%%%5
%%%%%%%%%%%%%%%%%%%%%%%%%%%%%%%%%%%%%%

 \begin{figure}
 \includegraphics[scale=1.5]{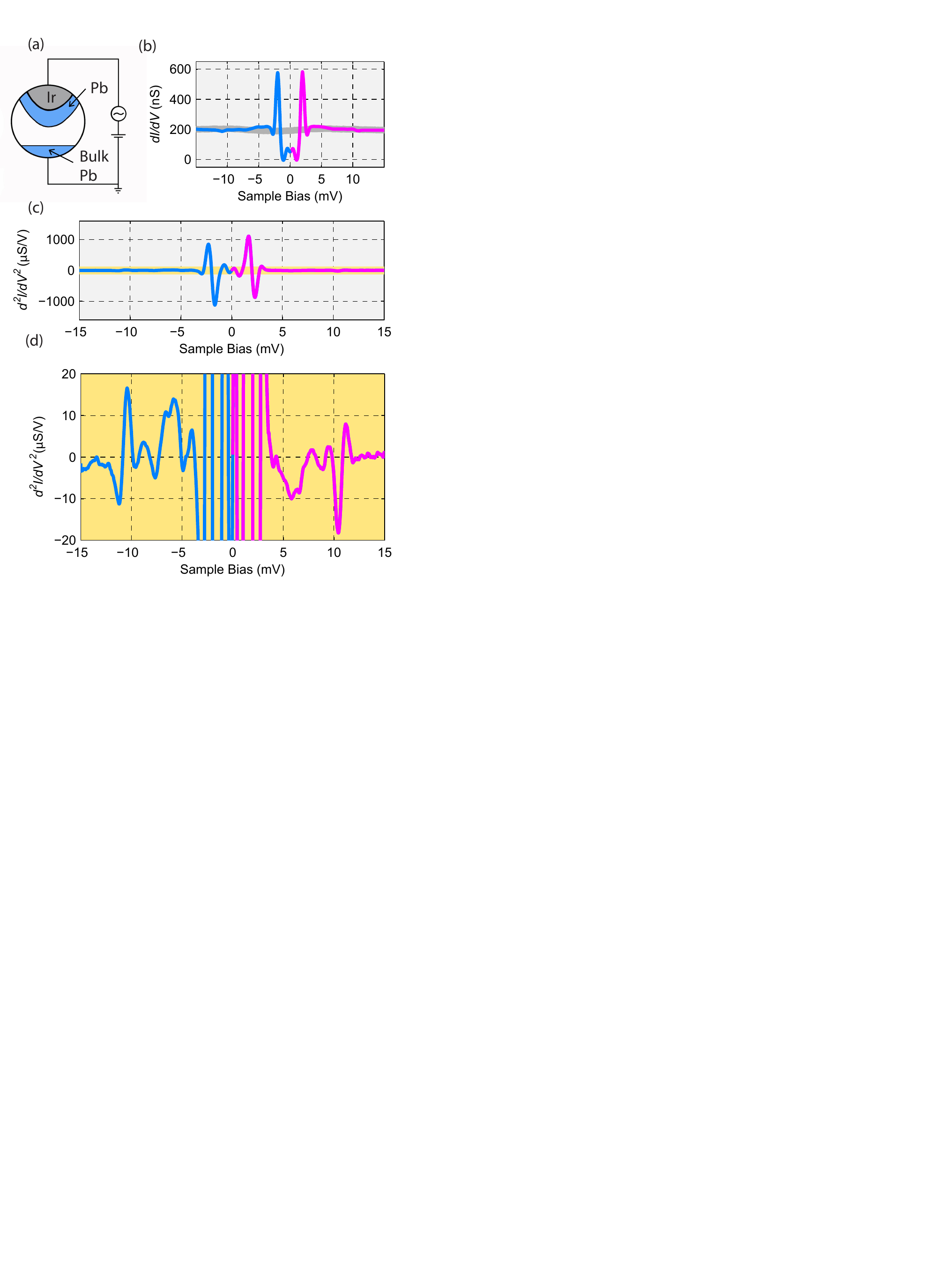}

 \caption{(a)  A diagram of the experimental S-Vac-S setup.   (b) Raw $dI/dV$
spectrum at a single location.  The negative sample biases, corresponding to
occupied electronic states, are colored in blue while the positive biases,
corresponding to unoccupied electronic states, are colored in magenta. The
spectrum in gray is the normal state spectrum acquired in a magnetic field $B=1$ T. (c) $d^2I/dV^2$ by
numerical differentiation of the data in (b).  The yellow band highlights the range of interest which we zoom in on in (d), showing high-resolution $d^2I/dV^2$ acquired by HHS.  }
 
 \label{fig:didvandd2idv2}
 
  \end{figure}	

%%%%%%%%%%%%%%%%%%%%%%%%%%%%%%%%%%%%%%%%%	
%%%%%%%%%%%%%%%%%%%%%%%%%%%%%%%%%%%%%%%%%%%%%%%%%%%%%%%%%%%%%%%%%%%%%%%%%%%%%%%%%%%		
	
Remarkably, despite the increasing prevalence of STS bosonic studies, 
no control experiment has ever been completed where vacuum-based STS is performed on a known conventional  
system to measure and fully invert the bosonic substructure, verifying it 
against the known result.  This is especially important since certain details 
of the tunneling process are still under 
investigation \cite{ PhysRevB.31.805, PhysRevB.42.8841, PhysRevB.43.11612, pilgramandrice, PhysRevLett.102.037001, PhysRevB.81.214512, PhysRevB.84.155414}.
Though conventional superconductors are better understood than their
unconventional counterparts, acquiring high-resolution STS spectra on
conventional superconductors is more challenging.  This
difficulty stems from significantly smaller gap sizes (an order of magnitude) 
which lead to more stringent requirements on noise minimization for 
resolution comparable to that of high-$T_c$ systems. 
(These requirements become even more stringent
when recording a higher harmonic of the modulation voltage to acquire
$d^2I/dV^2$).  There have been numerous prior results for vacuum STS on
conventional superconductors but these results have either ignored the bosonic 
substructure of the spectra while focusing on in-gap states, or not
possessed the required energy resolution and noise levels for clean
spectroscopy \cite{heinrichnbnb, xuepb,nbtipdavis}. 
In light of this, as diagrammed in Fig.~1(a) we studied STS on 
the archetypal elemental superconductor Pb, employing a superconducting Pb tip 
for feature sharpness and to construct a true vacuum analogue (S-Vac-S) to the
M\&R device study (S-I-S) \cite{PhysRevLett.14.108}.  (The reasons for this specific
material choice, as well as the detailed experimental method, can be found in the
Supplemental Material [SM]). We believe this to be the first inversion study of the phonon spectrum of a conventional superconductor utilizing ultra-high vacuum STS in true tunneling
mode.

To verify the superconducting nature of the sample 
and tip we measured a $dI/dV$ spectrum using the standard lock-in technique
[Fig.~\ref{fig:didvandd2idv2}(b)].  The spectrum 
shows the familiar structure for tunneling between two inequivalent superconductors 
with tall coherence peaks located at $\Delta = \Delta_\text{tip}+\Delta_\text{sample}$ 
($\Delta_\text{sample}>\Delta_\text{tip}$) 
and two in-gap features at $\Delta^\prime = \Delta_\text{sample} - \Delta_\text{tip}$.  
From this we obtain $\Delta = 2.0 \pm 0.02$ mV and $\Delta^\prime = 0.42 \pm 0.02$ mV which 
corresponds to superconducting energy gaps $\Delta_\text{sample} = 1.21\pm 0.02$ mV and $\Delta_\text{tip} = 0.79\pm 0.02$ mV. 
The $\Delta_\text{sample}$ value agrees well with the expected bulk value and the $\Delta_\text{tip}$ 
value is consistent with other experimental gap measurements on few-monolayer Pb
systems \cite{sciencepbshih,PhysRevLett.102.207002}. 
 	  
Continuing outwards from the coherence peaks, sharp dips are observed 
in $dI/dV$ on both sides of the Fermi energy $\epsilon_\text{F}$ ($V=0$) at energy $E_\text{dip} = 2.72 \pm
0.02$ mV. This value is equal (within experimental error) to $\Delta_\text{sample} + 2\Delta_\text{tip}$, but this may be a coincidence.  
Following the large dips are several step like features that are well known to be 
associated with electron-boson coupling (phonons, in this case). 
These features are best analyzed by examining $d^2I/dV^2$ vs.\ $V$ [Fig.\ 1(c)] which can be directly related \cite{PhysRevLett.14.108} to the effective electron-boson coupling function $\alpha^2(\omega)$ and the boson density of states $F(\omega)$ for bosons of energy $\omega$.  
We will discuss the $dI/dV$ dips in detail later in the Letter and first focus on the  
bosonic features (which are reported in terms of $\Omega$ where $\Omega=\omega-\Delta$). 

%%%%%%%%%%%%%%%%%%%FIgure%%%%%%%%%%%%%%%%%%%%%%%%%

 \begin{figure}
 \includegraphics[scale=1.5]{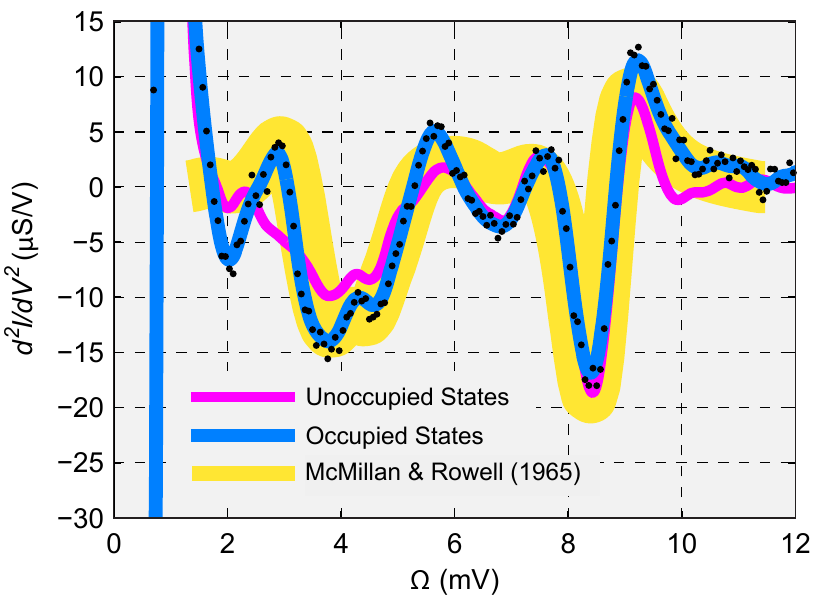}%
 \caption{  Smoothed $d^2I/dV^2$ vs.~$\Omega$ comparing the occupied and unoccupied spectra from this experiment with the M\&R device result \cite{PhysRevLett.14.108}. (Occupied curve is $-d^2I/dV^2$). The M\&R data has been thermally broadened from the 0.8 K at which the data was recorded to 4.2 K to match this experiment (see [SM]).  The occupied electronic states spectrum is accompanied by the raw data (black circles).  } 
\label{fig:compared2idv2} 
 \end{figure}
%%%%%%%%%%%%%%%%%%%%%%%%%%%%%%%%%%%%%%%%%%%%%%%%%%%%%

The largest $d^2I/dV^2$ signal [Fig.\ 1(c)]  comes from the sharpness of the coherence peak
which is two orders of magnitude larger than the other peaks.  In order to resolve
the much weaker signal from the boson substructure it is necessary to tune the settings of the tunnel junction and lock-in preamplifier 
to zoom in specifically on these higher energy features.  To highlight these complex spectral features we utilize a second lock-in preamplifier tuned to twice the modulation frequency to directly resolve
$d^2I/dV^2$, which we will refer to as higher-harmonic spectroscopy (HHS) [Fig.\ 1(d)].   

 In order to compare this data with
the M\&R device data, the M\&R result has been
digitized, thermally broadened, and overlaid with this study's result (see [SM]).   
It can be seen that the STS based S-Vac-S
data is in excellent agreement with the M\&R device S-I-S study (Fig.~\ref{fig:compared2idv2}).
The differences between the two sets of data will be highlighted and further analyzed 
later in the Letter.

%%%%%%%%%%%%%%%%%%%%%%%%%%%%%%%%%%
\begin{figure}
 \includegraphics[scale=1.5]{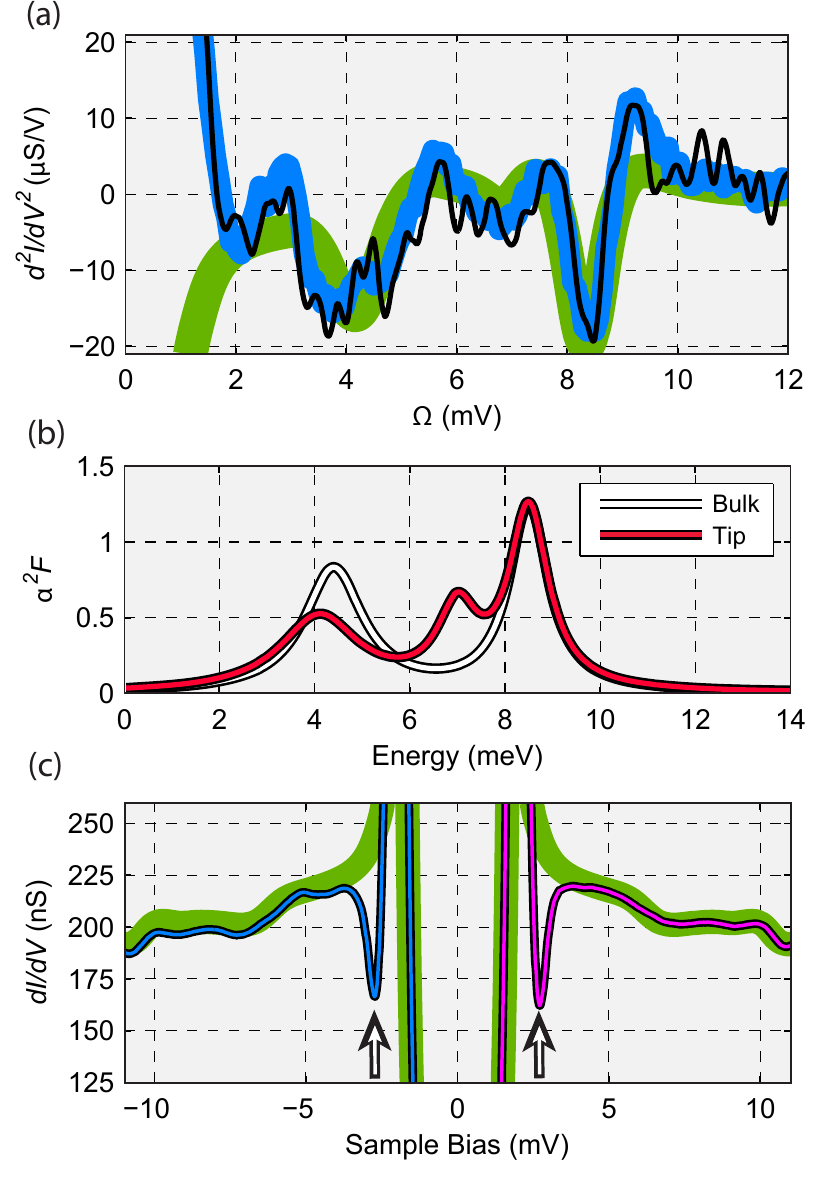}%
\caption{(a) Comparison of occupied $-d^2I/dV^2$ data by HHS experiment (blue), 
$-\frac{d}{dV}\left( \frac{dI}{dV}|_\text{SC} / \frac{dI}{dV}|_\text{N} \right)$ experimental data (black),  Eliashberg calculation based on model $\alpha^2F(\omega)$ (green). (b) Model $\alpha^2F(\omega)$ altered from the Lorentzian model of Scalapino \textit{et al.} \cite{Scalapino}. (c) Experimental $dI/dV$ curve overlaid with Eliashberg-based simulation (green).  Note the strong agreement with the exceptions of the large dips symmetric about $\epsilon_\text{F}$ (white arrows). } 
 \label{fig:specwithmag}
\end{figure}

 %%%%%%%%%%%%%%%%%%%%%%%%%%%%%%%

Unlike HHS, the M\&R method calls for taking $dI/dV$ in the
superconducting state, dividing it by $dI/dV$ in the normal state, and
differentiating the result.  This was necessary in order to 
divide out any spurious signals resulting from the insulating tunneling barrier which might
contribute inelastically to the tunneling process.  Since in our experimental 
setup the vacuum acts as a perfect insulating barrier, both methods should give similar results.  In order to substantiate this claim we repeat this experiment using the traditional M\&R method utilizing a 1-T magnetic field (the critical field of Pb is $\approx 0.080$ T \cite{RevModPhys.26.277}) and compare the results.  It can be seen that the
normal state curve is essentially flat when compared to the superconducting
state [Fig.~\ref{fig:didvandd2idv2}(b)], so it can be accurately  surmised that dividing $dI/dV|_\text{SC}$ by an approximately 
flat line will result in essentially the original curve.  For thoroughness we nevertheless
 complete the M\&R-prescribed recipe to obtain $d^2I/dV^2$ using the superconducting and normal state data of Fig.\ 1(b). The resultant
spectra is overlaid on the HHS result in Fig.~\ref{fig:specwithmag}(a).  It is evident
that the spectra are identical with the exception that the M\&R method is
much noisier.  This verifies that this normal-state division is not necessary
for S-Vac-S tunneling in this simple {\textquoteleft}control{\textquoteright}
case of tunneling between similar elemental superconductors. 
This is an important statement for future experiments.  Clearly it is
experimentally simpler to take one measurement and use raw output rather than
to take two measurements at different times, under different conditions, and
in combination with a numerical derivative combining the data sets.  We have
now shown that this efficient single HHS measurement is sufficient. (It is
important to note that this is
valid since the normal-state spectra is essentially flat.  In the specific case
of the hole-doped cuprates, where the normal state is highly non-trivial and
spatially dependent, such a division might still be necessary \cite{Pasupathy2008}).

Since our spectra is so similar to that of M\&R it should be 
straightforward to calculate the tunneling spectra 
with Eliashberg theory 
and match it to our experiment.  The tunneling current between two
superconductors separated by an insulating (or vacuum) barrier is given by 
%\begin{equation}\label{Eq:I}
$I(V) \propto \int_{0}^{V} d\omega N_{\text{SC},1}(\omega)N_{\text{SC},2}(\omega+V)[n_\text{F}(\omega) - n_\text{F}(\omega+V)]$
%\end{equation} 
where $V$ is the sample bias, $n_\text{F}(\omega)$ is the Fermi occupation 
number, and $N_{\text{SC},1}(\omega)$ and $N_{\text{SC},2}(\omega)$ are the density of states
(DOS) for the two respective superconductors.  Once $I(V)$ is known, the
differential conductance $dI/dV$ and the second derivative $d^2I/dV^2$ can
easily be found numerically. For a strong-coupling superconductor such as Pb, the DOS is given by 
%\begin{equation}\label{Eq:DOS}
$N_\text{SC}(\omega)/N_\text{N}(0) = \mathrm{Re}[\omega/\sqrt{\omega^2 - \Delta^2(\omega)}]$
%\end{equation}
, where the S (N)  subscript denotes the superconducting (normal) state, 
and $N_\text{N}(0)$ is the DOS at $\epsilon_\text{F}$.  
Here, $\Delta(\omega)$ is the complex frequency-dependent gap function given 
by the ratio of the anomalous self-energy and the renormalization function 
$\Delta(\omega) = \phi(\omega)/Z(\omega)$.  $Z(\omega)$ and $\phi(\omega)$ are 
obtained by solving the real-axis Eliashberg equations \cite{Carbotte, Scalapino} 
using $\alpha^2F(\omega)$ as an input and our calculations  
are therefore dependent on this choice. In this case we have two phonon spectra to 
consider, one for the bulk sample and one for the tip. For the former we follow 
Scalapino \textit{et al.} \cite{Scalapino}  and define $\alpha^2F$ with a sum of Lorentzians (for more details see [SM]). For the latter we modify this model with additional peaks to account for the thin-film nature and geometry of the superconducting tip where coupling can occur with phonon branches normally silent in bulk Pb.  Figure \ref{fig:specwithmag}  shows these 
model $\alpha^2F$ functions and the resulting $dI/dV$ and $d^2I/dV^2$ which agree with the 
experiment (acquired in the two previously described experimental methods).  
The strongest difference between the two are the strong 
dips immediately past the coherence peaks  in $dI/dV$ [indicated by the arrows in 
Fig.~\ref{fig:specwithmag}(c)].  These dips are also manifested in $d^2I/dV^2$ where the curves split at low energies.

The pronounced dips in $dI/dV$ are visually reminiscent of other tunneling studies---for example,
break junction tunneling experiments on hole-doped BSCCO-2212
\cite{PhysRevLett.80.153} where the authors claim a connection between 
the spectral dip shapes (often referred to as peak-dip-hump) and bosonic modes.  Similar peak-dip-hump 
structures have been seen numerous times in N-Vac-S standard STS studies
on hole-doped BSCCO-2212 as well.  Here we make no statement on the 
origin of these dips in the high-$T_c$ superconductors. We simply note that in our elemental
superconductor study very similarly shaped dips occur and, in this benchmark case, their absence in $\alpha^2 F(\omega)$ 
excludes those dips from being relevant to pairing. Similarly shaped dip features have previously been seen in STS S-Vac-S
junctions \cite{nbtipdavis,magneticmolonpb,multigapsc}, but none of
these studies give a compelling explanation for their existence. None of those studies though were able to truly differentiate that dip feature from other bosonic signals as in this work.  The most prevalent hypotheses for these features invoke proximity effect
where extra features in $d^2I/dV^2$ are observed in combined S-I-NS junctions
\cite{wolfzasaprox}, surface impurity considerations \cite{magneticmolonpb}, or $k$-selective tunneling processes in two-band superconductivity \cite{multigapsc}.  While a multiband model might not be appropriate for the case of Pb, the underlying idea of the $k$-selective tunneling might be relevant due to Pb's anisotropic superconducting gap \cite{PhysRevB.13.4738, PhysRev.153.513}.  In this experiment the direction of anisotropy for the tip might not align with the anisotropic axis of the bulk.  This alignment mismatch may relate to the similar dips in the break junction results between superconductors but does not elucidate peak-dip-hump behavior in standard STS studies with a normal metallic tip. 

The sandwich configuration of the M\&R device necessarily demanded that occupied and unoccupied states 
be equivalent since the definition of the ``sample'' electrode and ``tip'' electrode could always be re-designated due to identical geometry of the two electrodes. In this experiment, the tunneling electrodes are distinguishable allowing us to 
differentiate between the occupied and unoccupied states of the bulk sample. Due to the very 
weak particle-hole asymmetry in Pb, theory demands---and our simulations demonstrate---that no differences 
between the occupied and unoccupied sides should exist. In our measured
spectra the two sides are indeed mostly similar, but variations are in fact present. 
This particle-hole asymmetry does not have an obvious explanation.  Immediately, we can discount the notion 
of this being an artifact of the tip structure.  While it is possible for the electronic structure of the tip to add
slope information or distort the measured spectrum through convolution, there is no present theory to account for the tip subtly modifying the phonon structure on only one side of  $\epsilon_\text{F}$.  To explore these differences more thoroughly, both occupied and unoccupied $d^2I/dV^2$ spectra were fit with Gaussian peaks.  Two different fitting schemes were
explored.  As a starting point in fitting, the individual peak positions were
constrained to be identical for both the occupied and unoccupied states.  The experimental asymmetry could be captured by either introducing an extra peak at
$\Omega=2.87$ meV in the unoccupied side or by allowing the FWHM of the dual
peak structure near 4 meV to have different values for the occupied and
unoccupied sides (see data in Fig.\ 2).  Neither of these fits have a physical basis 
within Eliashberg theory, however---extra 
phonon structure cannot exist on only one side of $\epsilon_\text{F}$, nor is there
a physical reason for the broadening of the phonon peaks to have different
values on alternate sides of $\epsilon_\text{F}$. Even more puzzling is the fact
that only certain phonon modes seemed to be broadened asymmetrically about $\epsilon_\text{F}$ 
while the other modes have broadening symmetric about $\epsilon_\text{F}$.  

We now advance from the forecasted model of $\alpha^2 F(\omega)$ to a full inversion of the experimental data to calculate $\alpha^2 F(\omega)$.  Since the Eliashberg-based inversions are not compatible with occupied/unoccupied asymmetry calculations, this asymmetry can only be explored by treating the occupied and unoccupied sides of the spectra as independent data sets.  Another complexity of this calculation is the question of how to treat $\alpha^2 F(\omega)$ in terms of the sample superconductor and tip superconductor.  Since the inversion process is based on an iterative process to bring the calculated $d^2I/dV^2$ progressively closer to the experimental $d^2I/dV^2$, leaving both
the tip and the sample as separate fitted spectra would result in an infinite number of different solutions to $\alpha^2 F(\omega)$.  The most impartial way then to perform the 
inversion is  to constrain the magnitudes of $\alpha^2 F(\omega)_\text{tip}$ and
$\alpha^2 F(\omega)_\text{sample}$ in a fixed ratio determined by their respective $\Delta$ values
while allowing their linked overall shape to vary. (For a thorough description of the fitting procedure see [SM]).

%%%%%%%%%%%%%%%%%%%%%%%%%%%%%%%

 \begin{figure}[!]
 \includegraphics[scale=1.5]{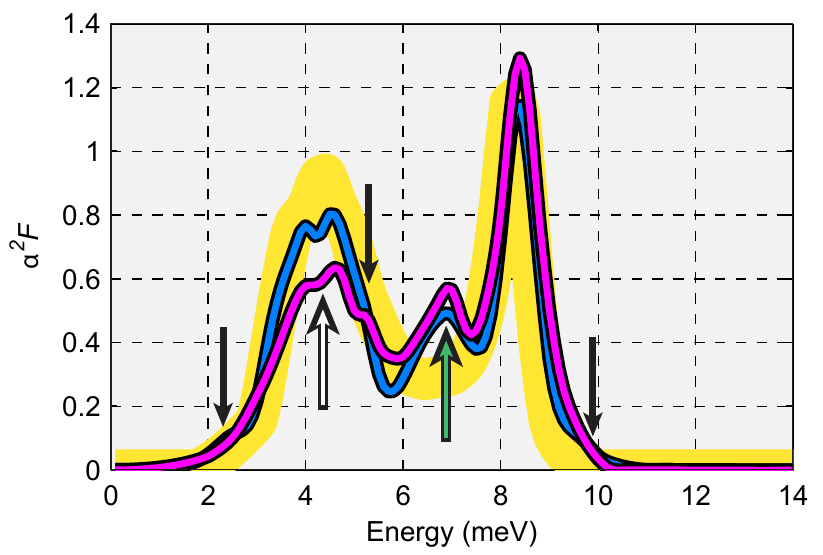}%
 \caption{Calculated $\alpha^2F(\omega)$ from spectral inversion of
 $d^2I/dV^2$ treating the occupied (blue) and unoccupied (magenta) electronic
 states as separate spectra. The yellow curve is the device $\alpha^2F(\omega)$ from
 M\&R \cite{PhysRevLett.14.108}.  Arrows mark features discussed in the text.   } 
\label{fig:invertedalpha} 
 \end{figure}
  
%%%%%%%end figure section

Fig.~\ref{fig:invertedalpha} shows the inverted $\alpha^2F(\omega)_\text{occupied}$ and $\alpha^2
F(\omega)_\text{unoccupied}$ and its general agreement with the M\&R result.  Using
the inverted $\alpha^2F(\omega)$ spectrum we now have the information necessary
to extract the dimensionless coupling strength  
$\lambda = \int_0^\infty { 2 \alpha^2F(\omega)d\omega}/\omega$
from our experimental data.
The occupied and unoccupied curves give $\lambda_\text{occupied}$ = 1.38 and
$\lambda_\text{unoccupied}$ = 1.34 with $\lambda_\text{average} = 1.36$ compared to the
M\&R result of $\lambda_\text{device}=1.33$ \cite{PhysRevLett.14.108}.  (From a
particle-hole symmetric theory it is again unphysical to quote that
the coupling strength and thus the $T_c$ is variable for the occupied and
unoccupied sides of the spectrum, but from an experimental point of view  we
have two valid data sets to be used to obtain $\lambda$ values).

The two strongest differences between this experiment's calculated $\alpha^2
F(\omega)$ and the M\&R result are the additional spectral weight around 7 meV
(Fig.~\ref{fig:invertedalpha}, green arrow) and the splitting of the first peak
(Fig.~\ref{fig:invertedalpha}, white arrow).  
These can easily be explained by the thin-film nature of the Pb
tip.  In a first-principles phonon simulation \cite{PhysRevB.81.214519} it was
seen that in two-monolayer films the surface lifted the degeneracy of the
transverse mode resulting in distinct
transverse peaks.  It was also seen that increased electron density on the
surface resulted in stronger Pb-Pb bonds, stiffening the longitudinal peak
which accounts for the added spectral weight seen around 7 meV. (The splitting
of the transverse modes can also be explained by enhanced spin-orbit
coupling \cite{PhysRevB.81.174527}, but this option does not explain the weight at 7
meV).

This then takes us to the differences between the occupied and unoccupied
inversions of $\alpha^2 F(\omega)$.  The black arrows in
Fig.~\ref{fig:invertedalpha} highlight these differences. The largest difference
is an additional peak in the transverse mode (black arrow near 5 meV).  This peak sits roughly in the middle of the range of 3--7 meV in which the occupied and unoccupied spectral data display the most significant contrast (Fig.\ 2), and 
is near the 5-meV energy where some shear vertical surface phonons have
been measured at energies that have been potentially softened by spin-orbit
coupling \cite{0953-8984-24-10-104004}.  This does not offer any explanation as
to why this would cause this phonon signal to be different when measured on the
occupied vs.\ unoccupied side.  The other difference between the occupied and
unoccupied inversions is that $\alpha^2 F(\omega)_\text{occupied}$ has two very small
peaks, one near 2 meV and one near 10 meV.  Since the coupling strength scales
inversely with $\omega$, the 2-meV peak is of far greater importance.  A peak
near this 2 meV energy is also seen both in the occupied and unoccupied spectra
of $d^2I/dV^2$ (although at different strengths).  This peak energy is very
close to the Kohn anomaly in the
transverse acoustic phonon at $q = (0.26, 0.26, 0)$ measured at energy $E=2.32$ meV
recently by inelastic neutron scattering \cite{kohnanomalies}. It had been
suggested that such a Kohn anomaly is caused by spin-orbit coupling or many-body interactions and has a correlation to the size of the superconducting gap
$2\Delta$.  This correlation though is controversial and may be
coincidental \cite{PhysRevB.84.174523}.  Nothing about the way we understand
these anomalies  lends a suitable explanation as to why they would appear
stronger or weaker (or not exist at all) in the occupied vs.\ unoccupied
spectra.  Theories that include electron-hole asymmetries \cite{Hirsch200621}
may need to be considered.

This work was supported by the U.S.~Department of Energy, Office
of Basic Energy Sciences, Division of Materials Sciences and Engineering, under
contract DE-AC02-76SF00515. F.~N. acknowledges support from the Karl A. Van Bibber 
Fellowship and S.~J. acknowledges support from the Foundation for Fundamental Research on Matter
(FOM, The Netherlands)  We thank A. V. Balatsky and M. R. Beasley for
discussions and comments. 

%\bibliographystyle{h-physrev3}	% (uses file "plain.bst")
%\bibliography{pbbib}		% expects file "myrefs.bib"

\end{document}